\documentclass[twocolumn,twocolappendix]{aastex7}

\usepackage{booktabs}
\usepackage{amsmath}
\usepackage{bm}
\usepackage{siunitx}

\begin{document}

\title{Crustal Quakes Spark Magnetospheric Blasts:\\ 
Imprints of Realistic Magnetar Crust Oscillations on the Fast Radio Burst Signal} 

\author[orcid=0009-0006-2620-0328,gname=Louis,sname=Burnaz]{Louis Burnaz}
\affiliation{TAPIR, Mailcode 350-17, California Institute of Technology, Pasadena, CA 91125, USA}
\email[show]{lburnaz@caltech.edu}
\author[orcid=0000-0002-0491-1210,gname=Elias Roland,sname=Most]{Elias R. Most}
\affiliation{TAPIR, Mailcode 350-17, California Institute of Technology, Pasadena, CA 91125, USA}
\affiliation{Walter Burke Institute for Theoretical Physics, California Institute of Technology, Pasadena, CA 91125, USA}
\email{emost@caltech.edu}  

\author[orcid=0000-0002-9711-9424, gname=Ashley,sname=Bransgrove]{Ashley Bransgrove}
\affiliation{Princeton Center for Theoretical Science, Princeton University, Princeton, NJ 08544, USA}
\affiliation{Department of Astrophysical Sciences, Princeton University, Princeton, NJ 08544, USA}
\email{abransgrove@princeton.edu}

\begin{abstract}

Many transients believed to originate from magnetars are thought to be triggered by crustal activity, which feeds back on the surrounding magnetosphere. These perturbations, through a variety of proposed mechanisms, can convert a fraction of the magnetic energy stored in the magnetosphere, as well as the energy injected by crustal activity itself into electromagnetic emission, including X-ray bursts and fast radio bursts. Here we provide a first glimpse of this process by coupling magneto-elastic dynamics simulations of the crust to fully three-dimensional relativistic resistive force-free electrodynamic simulations of the magnetosphere. Our simulations demonstrate that  the elastodynamical motions of the surface launch a series of fast magnetosonic and Alfv\'en waves into the magnetosphere. These waves rapidly enter a nonlinear regime, ultimately giving rise to a wide range of phenomena, including monster shock formation, relativistic blast waves, trapped Alfv\'en waves, nonlinear Alfv\'en wave ejecta, and transient equatorial current sheets interacting with these waves. After the initial nonlinear phase, the magnetosphere is partially combed out, resembling a strongly perturbed split monopole configuration. Our results can offer hints and potential constraints on fast radio burst emission mechanisms, in particular for hyperactive repeating sources, by placing tight bounds on energy conversion efficiency, and possible quasi-periodic imprints on magnetospheric waves by elastic oscillations of the crust.

\end{abstract}


\section{Introduction} 

Magnetars are young, highly magnetized neutron stars that exhibit intense transient electromagnetic (EM) activity, primarily in the X-ray band (see, e.g., \citealt{Kaspi:2017m} for a recent review). This activity encompasses short X-ray bursts (sXRBs), large-scale X-ray outbursts, and giant flares in the X-ray and soft $\gamma$-ray regimes. Magnetars have also been proposed as potential sources of fast radio bursts (FRBs), a hypothesis that gained strong support following the detection of the galactic FRB200428, which was unambiguously associated with the soft gamma repeater SGR 1935+2154 \citep{Bochenek:2020frb}.

Prevailing models attribute these emissions to the dissipation of the magnetar’s immense magnetic field (see, e.g., \citealt{Zhang:2022uzl} for a recent review). The associated static magnetic energy is typically converted into dynamic EM energy through complex and nonlinear mechanisms, involving magnetospheric perturbations (MPs), such as magnetospheric waves, as supported by various theoretical and numerical studies. For instance, some models focus on Alfv\'en wave dissipation in the closed magnetosphere, where the waves can be trapped along closed magnetic field lines near the magnetic equator. Through an energy cascade to smaller scales, their energy is transferred to the pair plasma, heating it and leading to X-ray emission \citep{Thompson:1995sgr}. 

Other models explore the amplification (relative to the decaying background magnetic field) and nonlinear effects of these waves that produce more complex phenomena: nonlinear fast magnetosonic (often referred to simply as fast waves) pulses compressing and displacing current sheet beyond light cylinder eventually triggering a major reconnection event that leads to the emission of a radio burst (\citealt{Lyubarsky:2020frb}; \citealt{Mahlmann:2022eff}; \citealt{Wang:2023ifr}), nonlinear Alfv\'en ejecta that generate and drag equatorial current sheets at relativistic speed, inducing reconnection-powered X-ray emission (\citealt{Yuan:2020pea}; \citealt{Yuan:2022mbd}), 
charged-starved Alfv\'en waves \citep{Lu:2020nsg}, resonant Alfv\'en wave interactions \citep{Long:2024wef}, curvature radiation \citep{Kumar:2017yiq,Yang:2017tmb}, 
strong magnetized shocks dissipating their energy through coherent synchrotron emission to power radio or X-ray bursts (\citealt{Lyubarsky:2014mfe}; \citealt{Beloborodov:2017fmf}; \citealt{Metzger:2019frb}; \citealt{Beloborodov:2023mrs}; \citealt{Vanthieghem:2024van}), or potentially by inverse Compton processes \citep{Zhang:2021pfn,Qu:2024cjh}. 

The generation of such MPs is commonly attributed to starquakes—sudden crustal failures—that generate crustal perturbations (CPs), that eventually excite magnetospheric waves in the close magnetosphere (\citealt{Blaes:1989nsm}; \citealt{Thompson:2001gf1}; \citealt{Li:2016moa}) or magnetar flares—sudden magnetospheric reconnection—that release a tremendous amount of energy, directly perturbing the magnetosphere (\citealt{Lyubarsky:2020frb}; \citealt{Mahlmann:2022eff}). 

In the starquake model, the evolving magnetic stress in the crust can eventually exceed the crust’s elastic limit, causing it to fail
\citep{Horowitz:2009ya,Thompson:2016dkd}. Although the crust of a neutron star is virtually incompressible, it can still undergo shear deformation. The sudden failure of the crust triggers shear perturbations that propagate to the surface. Since the crust is an excellent conductor, the internal magnetic field is effectively frozen into the crustal material. As a result, these perturbations shake the footpoints of the magnetar's dipolar magnetic field at the surface, generating magnetospheric waves \citep{Bransgrove:2020qqv}. 

Several theoretical and numerical studies have investigated sXRB and FRB emission scenarios based on simplified toy models of CPs coupled to a magnetosphere (e.g., \citealt{Yuan:2022mbd}). Other work has presented simulations of crustal modes inside a neutron star \citep{Gabler:2010gg,Gabler:2010rp,Gabler:2014bza,Sagert:2022gwu}. The question of how any CP variability could potentially get imprinted on FRB emission \citep{Suvorov:2019rzz,Wadiasingh:2020fxy} has gained recent interest with the potential discovery of quasi-periodic oscillations in a hyperactive repeating FRB source \citep{Zhou:2025acx}.
Additionally, this source potentially places (tight) constraints on the energetics of the FRBs \citep{Zhang:2025qzn}, raising the question of the available energy budget, and connections to magnetar quakes \citep{2025ApJ...988...62L,2025ApJ...979L..42W}.

However, we caution that this could also be intrinsic to magnetospheric dynamics associated with radio emission in neutron stars \citep{Kramer:2023mga}. Potential energy losses in the initial propagation of these waves can be substantial \citep{Yuan:2022mbd,Most:2024eig}, which are in addition to potential damping of subsequent radio waves themselves \citep{Beloborodov:2023lxl,Beloborodov:2025oei,Sobacchi:2024yis,Lyutikov:2023flz}. Damping in part depends on the background magnetic field geometry, making it imperative to understand the feedback of quake dynamics onto the background magnetosphere. 

Motivated by this, in this work we address several questions:
First, we demonstrate the precise type, amount and nonlinear propagation of waves launched by a realistic magnetar quake. We then quantify the propagation efficiency of these waves as they nonlinearly propagate outwards and interact with/deform the background magnetosphere, placing constraints on the energy available to trigger subsequent emission. Second, we demonstrate the highly distorted, combed-out structure of the magnetosphere following a realistic magnetar quake.

We do so by performing the first global magnetospheric dynamics simulation (in the force-free electrodynamics (FFE) approximation) coupled to neutron star crustal evolution using linear magneto-elastic dynamics (see also \citealt{Bransgrove:2020qqv,qu2025threedimensionalnumericalsimulationsmagnetar}).

Throughout this paper, we adopt the Heaviside-Lorentz convention for the EM fields. Moreover, positions, lengths, times or durations are expressed using the magnetar radius $r_*$ and EM fields using the magnetar surface magnetic field strength $B_*$.

\section{Methods}

Our approach consists of two steps. First, we compute CPs by solving the three-dimensional magneto-elastic wave equation in the crust. We then  use the surface motions of the crust as a boundary condition for the magnetosphere and study the subsequent force-free dynamics. We discuss the methodology of these two components in turn.

CPs are computed using the theoretical and numerical approach of \citealt{Bransgrove:2020qqv,qu2025threedimensionalnumericalsimulationsmagnetar}, who use a spectral method to solve the incompressible,  magneto-elastic dynamics of a realistic neutron star crust. The density profile $\rho$ is calculated by solving the equations for general relativistic hydrostatic equilibrium using the SLy equation of state for a neutron star of mass $M_* =1.4~M_\odot$ and central density $\rho_c = 10^{15}$~g~cm$^{-3}$ \citep{Douchin_2001}. This gives a neutron star with radius $r_* = 11.66$~km and a crust of thickness $H = 860$~m. The solid crust extends from the crust-core interface at density $\rho_{\rm core} = 1.27 \times 10^{14}$~g~cm$^{-3}$ up to the solid-liquid boundary at $\rho_{\rm crys} \approx 10^{11}$~g~cm$^{-3}$, appropriate for a young magnetar with internal temperature $T\approx 10^9$~K. Our model includes a liquid ocean on top of the crust that extends from $\rho_{\rm crys}$ to $\rho\approx 10^6$~g~cm$^{-3}$, where the shear wave velocity approaches $v\approx c$ (the lower boundary of the magnetosphere). The shear modulus $\mu$ is calculated using the formula of \cite{Strohmeyer_1991}, and the internal magnetic field used in the crustal simulations has characteristic strength $B\approx 4\times 10^{14}$~G. We use radiation boundary conditions at the top and bottom of the crust to account for the loss of quake energy into the magnetosphere and the liquid core as MHD waves. The simulations in this work resolve the crustal dynamics with $(n_{\rm max},l_{\rm max},m_{\rm max}) = (50,100,100)$, for a total of $1$~million spectral modes. Further details of the numerical method can be found in \citet{Bransgrove:2020qqv}

We model the quake as the sudden yielding of a localized strain layer in the deep crust. Therefore, we set the initial crustal deformation as a shear layer localized in the deep crust, similar to \citet{Bransgrove:2020qqv}. The characteristic spatial scale of the deformation gradient is $\ell_0 \sim 10^4$~cm, similar to the local pressure scale-height, and the opening angle of the sheared region is $\Delta\theta\sim \pi/4$. The characteristic frequency of elastic waves excited by the deformation is $\omega_0\approx v_s /\ell_0 \sim  10^4$~rad~s$^{-1}$, where $v_s =\sqrt{\mu/\rho} \approx 10^8$~cm~s$^{-1}$ is the elastic wave speed in the deep crust. 

The CPs exhibit two main temporal behaviors: a periodic pattern characterized by intense peaks occurring every $\Delta t = 22\,r_*/c \sim 1$~ms, corresponding to the time required for the perturbation to complete a round trip through the crust after reflecting off the core $t_{\rm el} \approx 2H/v_{s}$, and an exponential decay with a characteristic timescale $\tau \approx 10$~ms, representing the draining of the quake energy into the core as Alfv\'en waves upon each reflection (see the left column of Figure \ref{fig:crustmagneto}) \citep{Levin:2006ck}. The quake also contains oscillations extending to much higher frequencies.

We conducted a simulation with the CPs epicenter located in the $x$–$z$ plane (the $z$–axis being the magnetic polar axis) at a colatitude $\theta = \pi/4$, i.e.\ halfway between the magnetic pole and the magnetic equator. At this latitude, as the poloidal and toroidal velocity components have similar amplitudes, we expect roughly equal amounts to be injected in fast magnetosonic and Alfv\'en waves \citep{qu2025threedimensionalnumericalsimulationsmagnetar}.

\begin{figure*}[ht!]
\plotone{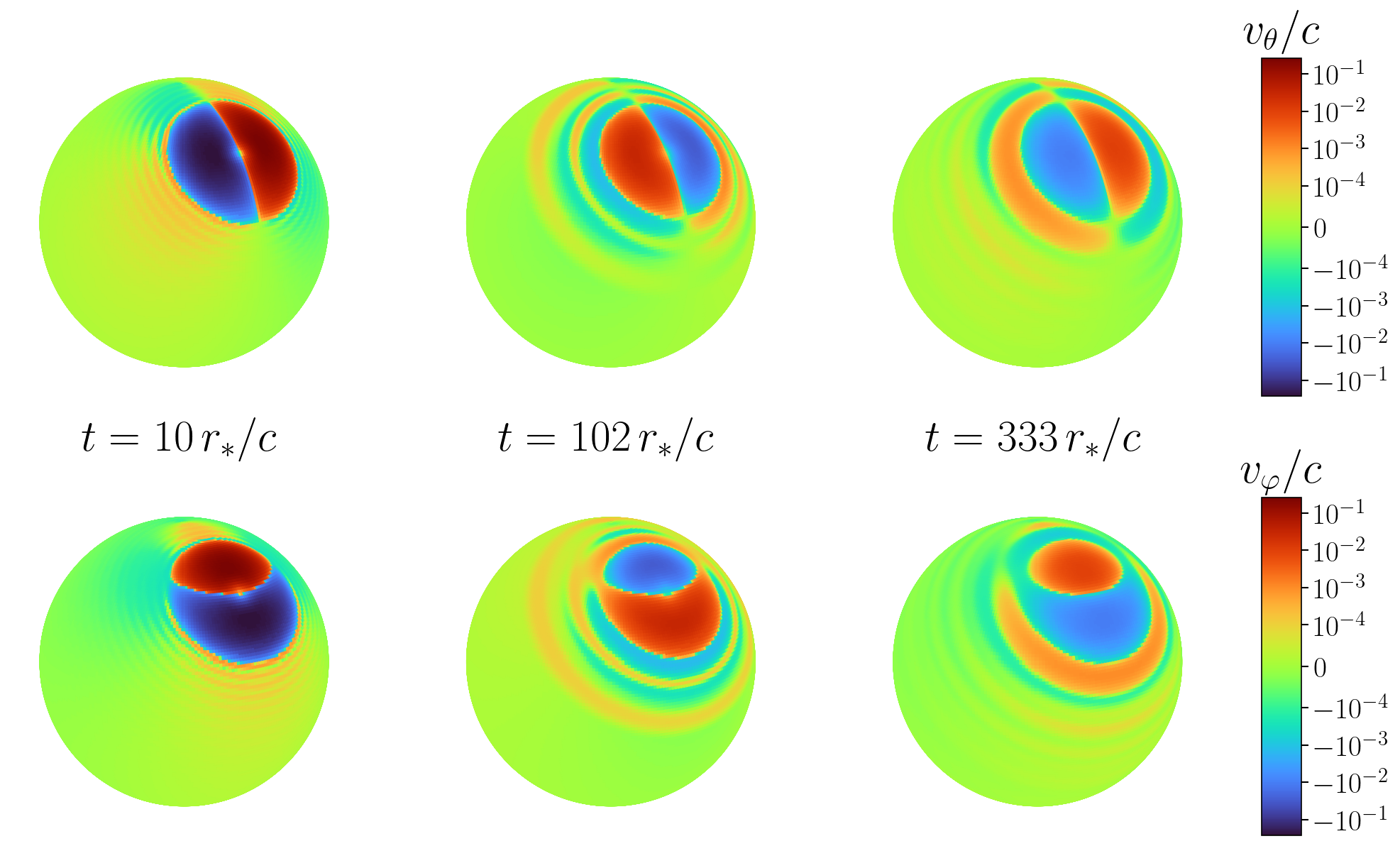}
\caption{Velocity profiles of the crustal perturbations {at the base of the magnetosphere}. The top row displays the polar velocity $v_\theta/c$, and the bottom row displays the azimuthal velocity $v_\varphi/c$, both shown at various times during the evolution of the perturbation. 
\label{fig:crustal_perturb}}
\end{figure*}

In order to ensure that the magnetospheric waves driven by these CPs exhibit nonlinear behavior within our simulation domain, we choose the amplitude of these waves to reach a maximum of $v/c \sim 0.1$ (Figure \ref{fig:crustal_perturb}).

{We model the dynamics of the magnetosphere using FFE (e.g., \citealt{Gruzinov:1999aza}), appropriate for the global dynamics of compact objects magnetosphere \citep{Goldreich:1969pe}. Such an approach is commonly taken when computing nonlinear wave dynamics over a large range in length scales, e.g., \citealt{Parfrey:2013gza,Yuan:2020pea,Yuan:2022mbd,Sharma:2023cxc,Mahlmann:2023ipm,Bernardi:2025awc} for magnetars, and  \citealt{Palenzuela:2013hu,Palenzuela:2013kra,Paschalidis:2013jsa,Most:2020ami,East:2021spd,Most:2022ayk,Most:2023unc,Most:2024eig,Mahlmann:2024keb,Skiathas:2025pnj} for binary systems.} Force-free electrodynamics imposes a certain number of constraints, namely
\begin{align}
    \mathbf{E}\cdot\mathbf{B}&=0\,,\label{eq:constraint1}\\
    B^2-E^2&>0\,.\label{eq:constraint2}
\end{align}
As a result, the force-free approximation will break down in charge-starved regions of the magnetosphere \citep{Chen:2020otx}, and cannot capture nonlinear wave dynamics, such as monster shock formation in the regime, where $E \rightarrow B$, which would be better modeled in a (radiative) magnetohydrodynamic (MHD) \citep{Beloborodov:2023mrs} {(see \citealt{Most:2024qgc,Kim:2024fuy} for global simulations)}  or collisionless kinetic approach \citep{Bernardi:2025gksa}. Consequently, in this initial work, we adopt a resistive force-free electrodynamics approach, which we can efficiently solve on a refined mesh structure on hundreds of graphics processing units (GPUs), see below for more details. We follow the formulation first introduced by \citet{Alic:2012asb,Palenzuela:2012my} and later extended by \citet{Most:2022epf}, in which the electric current density $\mathbf{j}$ is prescribed in a way that ensures the constraint \eqref{eq:constraint1} is satisfied. To this end, we introduce a conductivity $\sigma$, which enables energy dissipation in regions where the constraint \eqref{eq:constraint1} is violated:
\begin{equation}
    \mathbf{j}=c q \dfrac{\mathbf{E}\times\mathbf{B}}{B^2}+\sigma c\frac{\mathbf{E} \cdot \mathbf{B}}{B^2} \mathbf{B}\,,\label{eq:current}
\end{equation}
where $q$ is the electric charge density.
Rather than using a driving current for the magnetic dominance condition \eqref{eq:constraint2}, we manually limit the electric field to always be below a given threshold $E<0.999 B$. This shaving off of the electric field in emerging electric zones most closely resembles monster shock dynamics \citep{Beloborodov:2023mrs} observed in MHD simulations \citep{Most:2024qgc}.

\begin{figure*}
\centering
\includegraphics[width=0.9\textwidth]{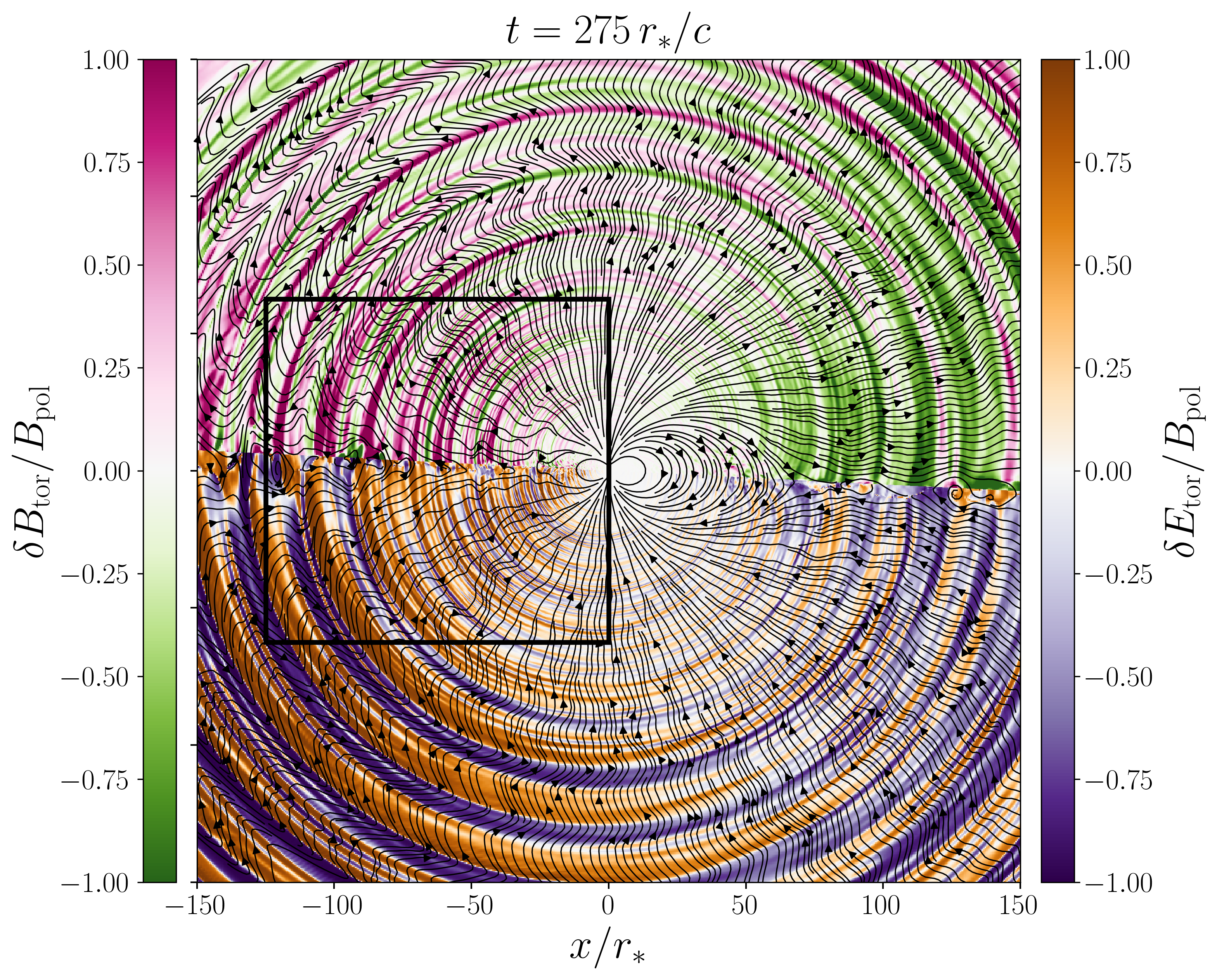}
\caption{{Magnetospheric state during a magnetar quake event. Shown is a} meridional $x$–$z$ plane slice at late times, $t = 275,r_*/c$, illustrating the perturbations of the toroidal magnetic field $\delta B_\mathrm{tor}$ (top part) and the toroidal electric field $\delta E_\mathrm{tor}$ (bottom part), normalized to the poloidal magnetic field $B_\mathrm{pol}$ in the full simulation{, corresponding to Alfv\'en and fast magnetosonic waves, respectively.} 
{We can see several cycles of wave injection, most of which nonlinearly steepen, combing out part of the initially dipolar magnetosphere.} The black square indicates the area shown in Figure \ref{fig:currentsheet}.}
\label{fig:dualfig}
\end{figure*}

\begin{figure*}
\centering
\includegraphics[width=\textwidth]{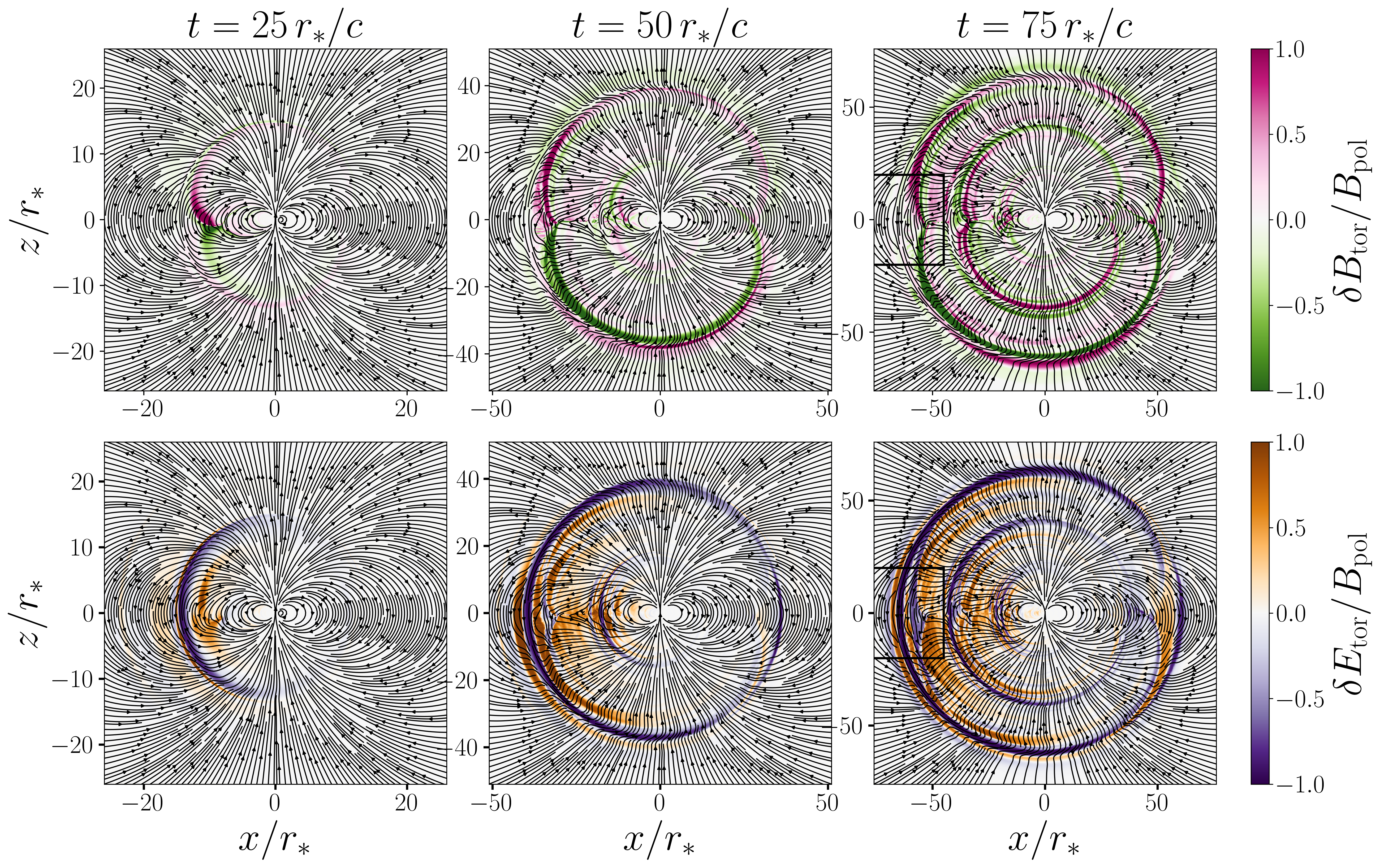}
\caption{Evolution of Alfv\'en and fast magnetosonic waves at early times, $t< 100 r_\ast/c$, where $r_\ast$ is the stellar radius. This is illustrated by the perturbations of the toroidal magnetic field $\delta B_\mathrm{tor}$ (top row) and the toroidal electric field $\delta E_\mathrm{tor}$ (bottom row), normalized to the poloidal magnetic field $B_\mathrm{pol}$, respectively. Both waves enter a nonlinear regime. In the case of Alfv\'en waves, forming a nonlinear Alfv\'en wave packet, opening the magnetosphere and ejecting it, forming equatorial current sheets. (Kilohertz) fast magnetosonic waves will ultimately enter a monster shock stage, indicated by the characteristic shape \citep{Beloborodov:2023mrs}, e.g., of the leading waves in the lower rightmost panel. The black square indicates the area shown in Figure \ref{fig:monstershock}.}
\label{fig:timeseries}
\end{figure*}

Numerically, we solve the relativistic FFE equations using a fourth order accurate conservative finite difference discretization using the ECHO scheme \citep{DelZanna:2007pk}, which is implemented on top of the AMReX mesh-refinement library \citep{amrex}. The code was originally used to study compact binary magnetospheres \citep{Most:2022epf}, including in full general relativity \citep{Most:2023unc,Most:2024eig}, and has recently been ported to GPUs (see \citealt{Most:2024eig} for details). The combination of using hundreds of GPUs with several levels of mesh refinement is crucial to overcome the large scale separation inherent to nonlinear wave dynamics we investigate here.

Once the CPs were obtained by solving the magneto-elastic dynamics equations, they were spatially regridded onto a high-resolution angular grid using cubic interpolation to match the resolution of our magnetosphere simulation. A linear interpolator was then used to inject the CPs as a dynamic inner boundary condition at the magnetar surface, by assuming an ideal electric field, ${\bf E}=- {\bf v}\times {\bf B}/c$.

For our simulations, we employed a cubic computational domain spanning $-400\le x_i/r_* \le 400$ for each coordinate $x_i \in \{x, y, z\}$, resolved using a uniform cartesian base grid of $768^3$ cells, corresponding to a base cell size of $\Delta_0/r_* = 1.04$. The base grid is progressively refined using nested cubic sub-domains, each spanning $-400/2^n\le x_i/r_* \le 400/2^n$ and a cell size reduced by a factor of $2^n$, i.e. $\Delta_n = \Delta_0 / 2^n$, for $n \in [1, 7]$. The highest refinement level reaches a cell size of $\Delta_7/r_* \approx 0.008$ within a cube of side length $6r_*$. The domain is centered on the magnetar, modeled as a perfectly conducting sphere of radius $r_*$, with a magnetic moment $\vec{\mu}$ along $z$-axis, inducing a dipole magnetic field $\vec{B}_\mathrm{dip}$. For improved numerical stability (primarily numerical noise) of the background solution, we found it advantageous to consider a rotating model, with large light cylinder, i.e., $r_*\vec{\Omega}/c=\vec{e}_z/200$ of the magnetar, aligned with its magnetic moment $\vec{\mu}$. 

All analyses are performed in a stationary asymptotic non-corotating frame, but we use a corotating coordinate system $(x, y, z)$ to decompose the different vector fields. No data is used outside the light cylinder $r_\mathrm{LC}/r_* = 200$.

\section{Results}

In this Section, we present the result of our crustal mode injection into a magnetar magnetosphere (Figure \ref{fig:dualfig}). We will begin with discussing the overall dynamics, before focusing on individual wave dynamics, and an overall energy budget quantification.

\begin{figure*}
\centering
\includegraphics[width=\textwidth]{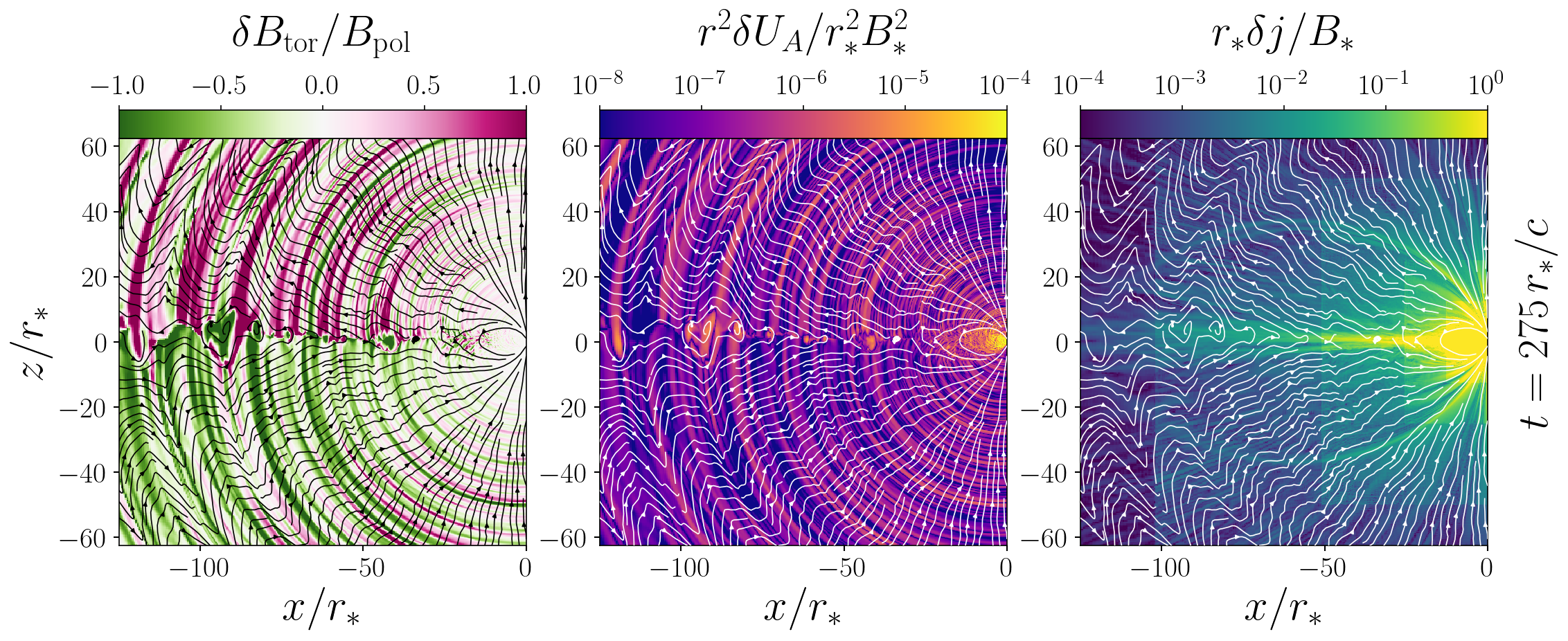}
\caption{Meridional $x$–$z$ plane slice at late time $t = 275\,r_*/c$, illustrating the perturbations of the toroidal magnetic field $\delta B_\mathrm{tor}$ normalized to the poloidal magnetic field $B_\mathrm{pol}$, the rescaled Alfv\'en waves energy density $r^2\delta U_A/r_*^2B_*^2$ and the rescaled current density $\delta j/r_*B_*$, where $r_\ast$ and $B_\ast$ are the radius and surface magnetic field strength of the neutron star.}
\label{fig:currentsheet}
\end{figure*}

In order to disentangle wave dynamics from the background properties, we have performed two simulations. One, without and one with the elastic modes injected. We then subtract the background (unperturbed) fields $\mathbf{X_0}\in\{\mathbf{B_0},\mathbf{E_0},\mathbf{j_0}\}$ from the perturbed fields $\mathbf{X}\in\{\mathbf{B},\mathbf{E},\mathbf{j}\}$ defining the perturbation fields as $\delta \mathbf{X} = \mathbf{X} - \mathbf{X}_0$.  Consequently, we define, e.g., the electromagnetic energy density of the perturbations by $\delta U = (\delta B^2 + \delta E^2)/2$ and the Poynting flux of the perturbations $\delta \mathbf{S} = \mathbf{E}\times\mathbf{B}-\mathbf{E}_0\times\mathbf{B}_0$. We further decompose each field $\mathbf{Y}$ into its toroidal component $Y_{\mathrm{tor}} = Y_\varphi$ and its poloidal amplitude $Y_{\mathrm{pol}} = \sqrt{Y_r^2 + Y_\theta^2}$, relative to the background dipole magnetic field and spin axis of the neutron star, which we take to be aligned.

As shown in Figure \ref{fig:timeseries}, the newly generated Alfv\'en and  fast waves are initially linear, with relative amplitudes $\delta B_\mathrm{tor}/B_\mathrm{pol} \sim \delta E_\mathrm{tor}/B_\mathrm{pol} \sim 0.01$, but they quickly amplify as they propagate outward through the decaying dipolar magnetic field, which remains largely unaffected. In particular, $\delta B/B_\mathrm{dip} \propto (r/r_*)^{3/2}$ for Alfv\'en waves \citep{Blaes:1989nsm,Yuan:2020pea,Yuan:2022mbd}, and $\delta E/B_\mathrm{dip} \propto (r/r_*)^{2}$ for the fast waves \citep{Beloborodov:2023mrs}. This amplification continues until reaching critical radii $r_\mathrm{\times, f}/r_*\approx 10$ for the fast waves $r_\mathrm{\times,A}/r_*\approx 20$ for the Alfv\'en ones, where the perturbations become fully nonlinear ($\delta B_\mathrm{tor}/B_\mathrm{pol} \sim \delta E_\mathrm{tor}/B_\mathrm{pol} \sim 1$). We present a full state of the magnetosphere after injection of a significant fraction of the CP in Figure \ref{fig:dualfig}. We can see that the resulting magnetosphere is no longer dipolar in shape, having been combed out by the strong nonlinear waves injected during multiple injection episodes. In order to understand the dynamics of this state and how it forms, we will in the following describe the individual wave dynamics.

\begin{figure}[b!]
\centering
\includegraphics[width=0.45\textwidth]{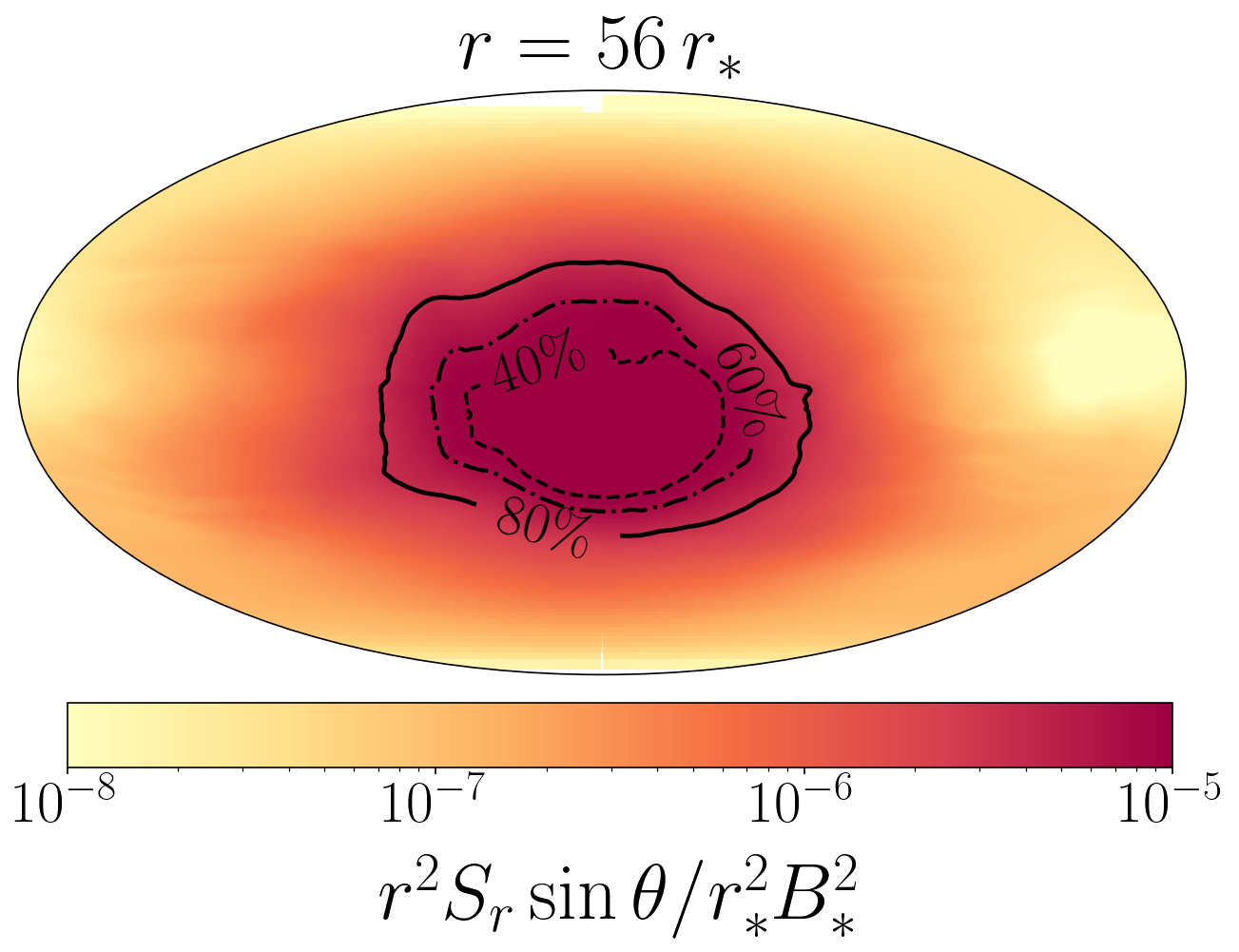}
\caption{Angular distribution of the radial energy flux associated with the first nonlinear fast pulse and nonlinear Alfv\'en ejecta, shown in a Mollweide projection. The Poynting flux on a spherical surface of radius $r = 56\,r_*$ was integrated over the time interval $ct/r_* \in [57, 77]$.}
\label{fig:skymap}
\end{figure}

\subsection{Alfv\'en wave and transient current sheet dynamics}\label{sec:Alfven}

At this stage, as shown in the first row of Figure \ref{fig:timeseries}, the equatorial nonlinear Alfv\'en waves, as they approach the apex of the background magnetic field loops along which they propagate, possess sufficient pressure to displace them, thereby initiating the opening of the closed magnetic loops. We stress that the opening of the background will generically consume a fraction of waves energy, which in the past has been found to be as high as $80\%$ \citep{Yuan:2020pea,Most:2024eig}. This is accompanied by the ejection of large plasmoids propagating outwards at relativistic speeds. The interaction of multiple such ejecta has been proposed to drive FRB emission via synchrotron maser processes \citep{Yuan:2020pea}. Additionally, this opening allows a fraction of later emitted Alfv\'en waves to escape radially as nonlinear Alfv\'en ejecta. However, as shown on the right-hand side of the second plot of Figure \ref{fig:currentsheet} showing Alfv\'en waves energy density $\delta U_A$, a significant proportion of Alfv\'en waves are not sufficiently amplified to become nonlinear, and therefore remain confined within closed magnetic field lines.

\begin{figure*}
\centering
\includegraphics[width=1\textwidth]{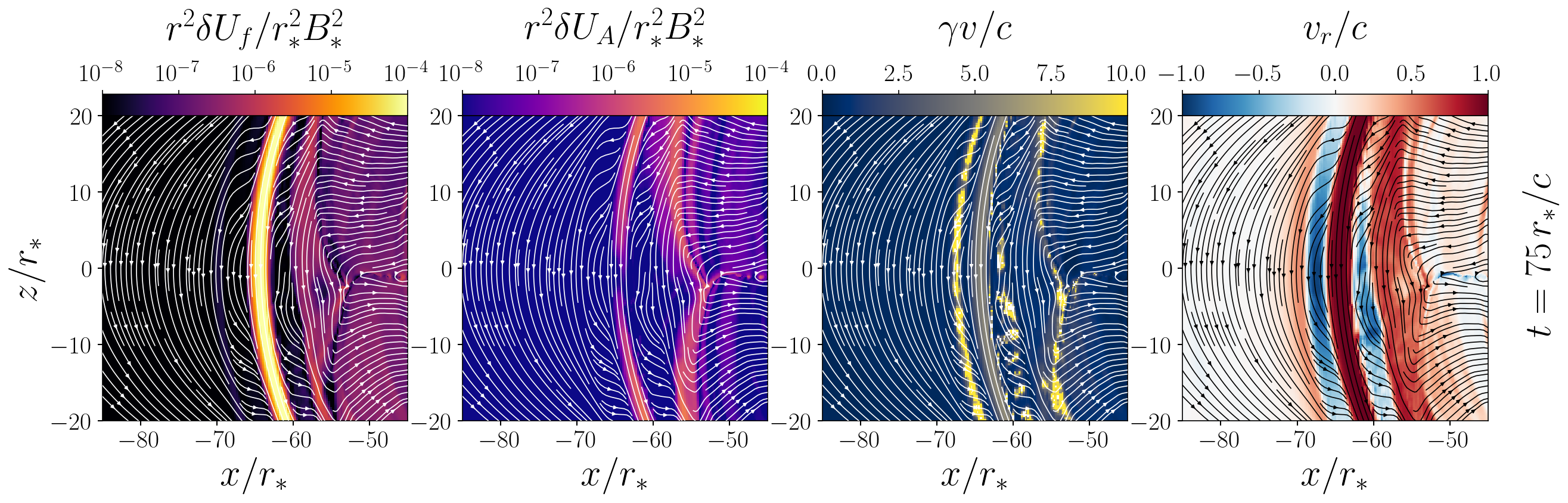}
\caption{Properties of the leading fast magnetosonic wave in the nonlinear regime. Meridional $x$–$z$ plane slice at time $t = 75\,r_*/c$, illustrating the rescaled magnetosonic and Alfv\'en wave energy densities $r^2\delta U_f/r_*^2B_*^2$ and $r^2\delta U_A/r_*^2B_*^2$, the relativistic drift speed amplitude $\gamma v/c$ and the radial drift speed  $v_r/c$. The wave is in a strong nonlinear regime, $v_r \simeq c$, and the geometry of the nonlinear zone resembles that of a monster shock \citep{Beloborodov:2023mrs,Bernardi:2025wht}.}
\label{fig:monstershock}
\end{figure*}

Trapped Alfv\'en waves are confined within $r < 12\,r_*$, and reach at most a mildly nonlinear regime with $\delta B_\mathrm{tor}/B_\mathrm{pol} \sim 0.4$. Thus, nonlinear mode conversion can occur, specifically interactions between Alfv\'en waves  and coupling with curved magnetic field lines \citep{Mahlmann:2024fwi,Bernardi:2024upq}, can lead to the generation of fast waves \citep{Yuan:2020eor}. These fast waves are no longer trapped and can propagate freely out of the closed magnetic field region. However, due to the complex spectral content of the CPs and MPs, it is not possible to unambiguously distinguish fast waves produced via these nonlinear interactions from those generated directly at the magnetar surface. Furthermore, the efficiency of these nonlinear mechanisms is too low to produce observable effects within the duration of our simulation, and no energy transfer from trapped Alfv\'en waves to nonlinear fast waves can be clearly identified. It is also possible that these nonlinear interactions between trapped Alfv\'en waves eventually trigger a turbulent energy cascade toward smaller scales \citep{Thompson:1998mer,2021JPlPh..87f9014T,2021JPlPh..87e9012R}, where resistive effects eventually dominate and dissipate the magnetic energy, thereby heating the magnetospheric plasma \citep{Zrake:2015hda}.

Nonlinear Alfv\'en ejecta drag the background dipolar magnetic field lines along their motion, forming and advecting transient equatorial current sheets \citep{Yuan:2020pea,Yuan:2022mbd}. These unstable current sheets eventually undergo magnetic reconnection, leading to the closure and detachment from magnetic field lines that are connected to the magnetar. These plasmoids are subsequently ejected and propagate outwards. Reconnection in these current sheets dissipates a significant fraction of the magnetic energy stored in the initially dipolar field at this place. In our simulation, the nearest unstable equatorial current sheets form around $r \sim 20\,r_*$, where the background magnetic field is approximately $10^{-4}\, B_*$. During reconnection, a fraction $\beta \sim 0.1$ of the local magnetic energy is dissipated through an energy cascade mediated by plasmoid collisions \citep{Hakobyan:2022kiy,Philippov:2019qud}. These collisions give rise to secondary current sheets, which themselves become unstable and reconnect, until resistive scale is reached and magnetic energy dissipated \citep{Guo:2015cua}, with the details depending strongly on the microphysics \citep{Hakobyan:2018fwg}. This may have implications for secondary X-ray emission from this process \citep{Yuan:2020pea}.

\subsection{Fast magnetosonic wave and monster shock dynamics}\label{sec:fast}

Having described the Alfv\'en wave dynamics, we now focus on the fast magnetosonic wave. Once fast magnetosonic waves enter the nonlinear regime, as shown in the second row of Figure \ref{fig:timeseries}, they strongly and temporarily distort the background poloidal magnetic field and continue to propagate radially outward. In detail, we identify two separate leading waves in Figure \ref{fig:timeseries}, a leading fast magnetosonic waves likely directly launched from the surface, and a second wave nonlinearly overlapping with the leading Alfv\'en wave, indicative of mode conversion \citep{Yuan:2020eor,Bernardi:2024upq}, especially as the background magnetosphere is opened up \citep{Yuan:2020pea,Yuan:2022mbd}. In this case, it seems that its magnetic field perturbation prevents the formation of an electric zone ($E^2 \gtrsim B^2$), indicative in our force-free approach of monster shock formation \citep{Beloborodov:2023mrs,Most:2024qgc}, limiting dissipation of the fast wave. However, we caution that our simulation does not correctly capture the MHD aspects of shock formation. 

In contrast, the leading fast wave appears to be unstable to monster shock formation, and is rapidly damped by our monster shock subgrid model, shaving off excess electric field. They form nonlinear low-frequency fast pulses with a typical width of $w/r_*=1.5$, a distance between pulses of $d/r_*=22$ and amplitude of $\delta E/B_*\sim10^{-2}\,r_*/r$. Figure \ref{fig:skymap} shows the typical angular distribution of energy of the first nonlinear fast pulse and nonlinear Alfv\'en ejecta, with 80\% of the energy contained in a zone of solid angle $\Omega/4\pi \sim 0.13$, indicating that the first MPs are tightly collimated.

We also remark that the scenario we find can naturally lead to fast wave collisions with the transient plasmoids in equatorial current sheet formed and advected by nonlinear Alfv\'en waves (Section \ref{sec:Alfven}). In the case of finite magnetization ($\sigma= v_A / \sqrt{c^2-v_A^2}$), Alfv\'en and fast magnetosonic group speeds are in general different ($v_A\neq c_{ms}$), so fast waves may eventually catch up to transient plasmoids, potentially causing local compression of the parts of the secondary sheet at an angle with the fast waves (see, e.g., the $x < -100 r_\ast$ regions Figure \ref{fig:currentsheet}), similar to what has been found, e.g., in simulations of merging neutron star binaries \citep{Most:2022ayk}.
This has been argued to lead to high frequency wave emission from collision of plasmoids in the sheet \citep{Lyubarsky:2020frb,Mahlmann:2022eff}, though at reduced energy efficiency compared to the other channels we discuss.\\

\begin{figure*}
\centering
\includegraphics[width=0.9\textwidth]{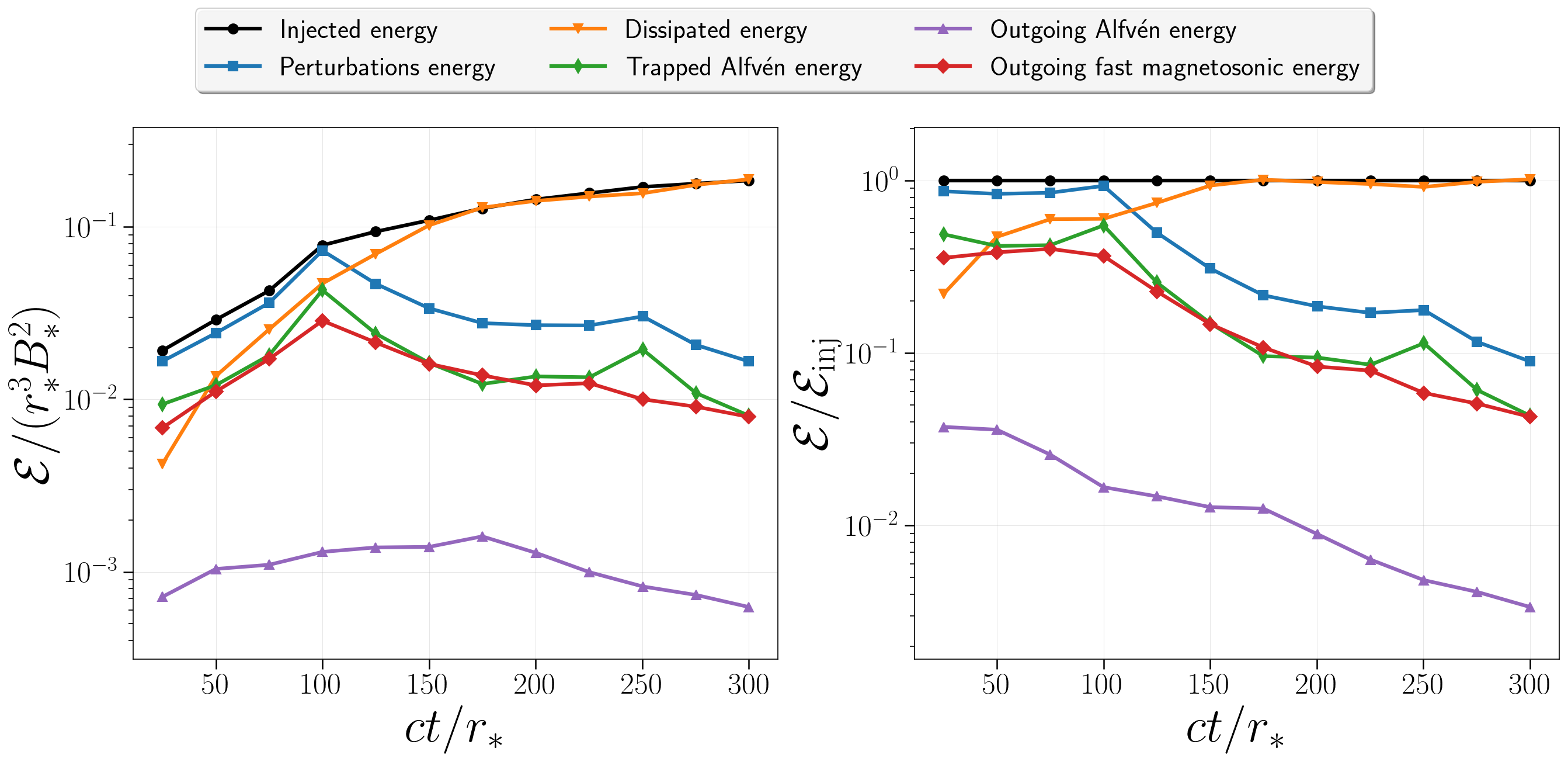}
\caption{Time evolution of different energy components, $\mathcal{E}$, within $r < 100\,r_*$. Here $r_\ast$ and $B_\ast$ are the stellar radius and magnetic field strength respectively. The left panel shows the various energy components, while the right panel presents their proportions relative to the injected energy. From $t = 100\,r_*/c$, the first perturbations escape the integration domain. At this stage the magnetosphere partially transitions into a split monopole configuration, making it more difficult to define (linear) wave perturbations as done here.}
\label{fig:energyproportion}
\end{figure*}

We now quantify some properties of the monster shock-like features. Figure \ref{fig:monstershock} illustrates an equatorial region centered on the first fast pulse, highlighting the development of several nonlinear electric zones characterized by $E^2 \lesssim B^2$ (limited by our monster shock subgrid model), or equivalently, regions where $\gamma v/c \gtrsim 1$, since $\gamma v/c = \sqrt{E^2/(B^2 - E^2)}$. Zones reaching $\gamma v/c \geq 10$ undergo significant dissipation and, at this stage, have already lost most of their initial energy, {due to limiting of the electric field in near electric zones.} The first fast pulse corresponds to the $\gamma v/c\sim5$ zone, featuring a strong radial outflow ($v_r/c\sim1$) preceded by an inflowing layer ($v_r/c<0$), which is the ideal precursor to a fast shock. By comparing the locations of the fast magnetosonic, $\delta U_f$, with that of the Alfv\'en wave, $\delta U_A$, energy densities we confirm our earlier picture of Alfv\'en wave-fast wave conversion as the background dipole magnetic field is opened up.
When looking at the angular morphology, the region shows a very strong resemblance of the angular shape of the monster shock region \citep{Beloborodov:2023mrs}, as seen in MHD \citep{Most:2024qgc} and kinetic \citep{Bernardi:2025wht} simulations. Overall, this leads us to believe that the fast wave   we see is consistent with the (fully expected) monster shock scenario \citep{Chen:2022yci,Beloborodov:2023mrs}.

\subsection{Energetics and quasi-normal modes}

In Figure \ref{fig:energyproportion}, we display the different energy components $\mathcal{E}$ derived from our simulation. To quantify the energetics of the MPs (relative to the initial dipole background) and of its Alfv\'enic and fast magnetosonic components, we computed the energies $\mathcal{E}_\mathrm{per}$, $\mathcal{E}_A$ and $\mathcal{E}_f$ by integrating over a spherical domain extending from $r/r_* = 1$ to $100$ respectively: The total perturbation energy density $\delta U = (\delta B^2 + \delta E^2)/2$, the Alfv\'enic energy density $\delta U_A = (\delta B_\mathrm{tor}^2 + \delta E_\mathrm{pol}^2)/2$, and the fast magnetosonic energy density $\delta U_f = (\delta B_\mathrm{pol}^2 + \delta E_\mathrm{tor}^2)/2$.

To distinguish between trapped and outgoing Alfv\'en components, we integrated $\delta U_A$ over two radial ranges: from $r/r_* = 1$ to $12$ for the trapped component, and from $r/r_* = 12$ to $100$ for the outgoing component.

\begin{figure*}
\centering
\includegraphics[width=0.9\textwidth]{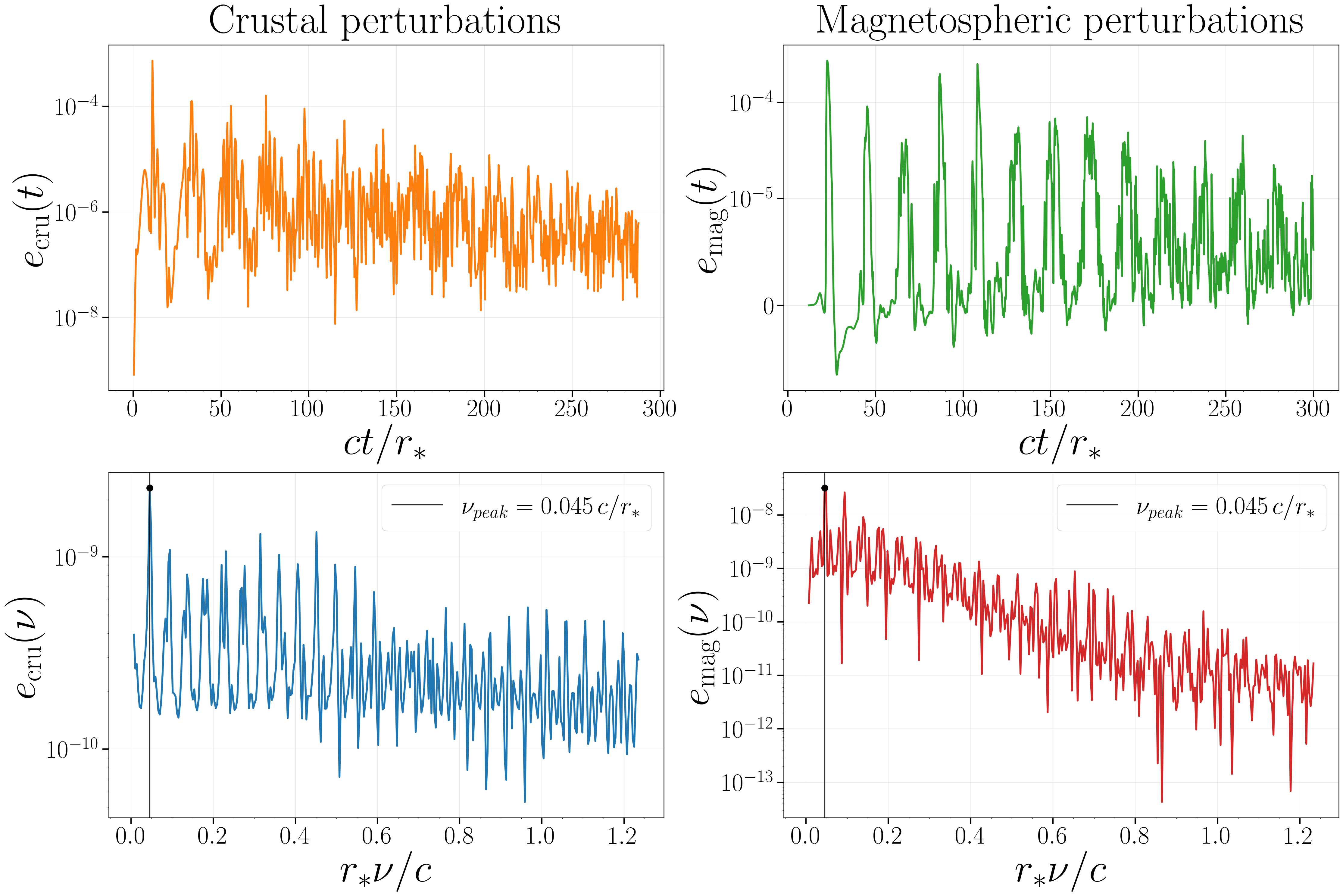}
\caption{Time evolution (top row) and corresponding frequency spectrum (bottom row) of {crustal surface motion} (left column) and magnetospheric (right column) perturbations. We {quantify surface motion via an effective normalized surface kinetic energy} $e_\mathrm{cru}=\frac{1}{2 c^2}\int(v_\theta^2+v_\varphi^2)\sin{\theta}\mathrm{d}\theta\mathrm{d}\varphi$ and {magnetospheric perturbations via the normalized electromagnetic energy} $e_\mathrm{mag}=\frac{1}{ cr_*^2B_*^2}\int r^2 S_r\sin{\theta}\mathrm{d}\theta\mathrm{d}\varphi$ where $S_r$ is the radial component of Poynting vector at $r=12\,r_*$, {with $r_\ast$ and $B_\ast$ being the stellar radius and surface magnetic field strength, respectively.} %
}
\label{fig:crustmagneto}
\end{figure*}

The injected energy $\mathcal{E}_\mathrm{inj}$ was calculated by integrating the incoming Poynting flux ($\delta \mathbf{S}=\mathbf{E}\times\mathbf{B}-\mathbf{E}_0\times\mathbf{B}_0$) across the inner boundary at $r = r_*$, both temporally and angularly, and subtracting the integrated outgoing Poynting flux at the outer boundary $r = 100\,r_*$.

Finally, the dissipated energy $\mathcal{E}_\mathrm{dis}$ was obtained by subtracting both the injected energy density and the unperturbed energy density $U_0 = (B_0^2 + E_0^2)/2$ from the total energy density $U = (B^2 + E^2)/2$, both integrated over the same spherical domain extending from $r/r_* = 1$ to $100$.

One important aspect to place engine driven constraints on magnetospheric dynamics, is the overall conversion efficiency, $\mathcal{E}/\mathcal{E}_{\rm inj}$, of the injected wave energy, $\mathcal{E}_{\rm inj}$, into escaping waves and nonlinear wave ejecta, as this would set the initial conditions for subsequent processes producing high-energy emission. As the waves propagate outwards, a sizable fraction of the initial wave energy is spent opening and deforming the background magnetic field into a partially split monopole magnetosphere.

As a result and as shown in Figure \ref{fig:energyproportion} only $7\pm 2\%$ of the Alfv\'en wave energy and $8\pm2\%$ of the outgoing energy (and so $4\pm 1\%$ of the total perturbations energy) is carried away by nonlinear Alfv\'en ejecta. During their radial propagation, these ejecta continue to slowly lose energy by displacing the background magnetic field lines. Trapped Alfv\'en waves contain around $52\pm 5\%$ of the total perturbations energy, On the other hand, nonlinear fast waves always have the largest share of outgoing energy  representing up to $44\pm4\%$ of the total perturbation energy. Part of this energy comes from mode conversion of nonlinear Alfv\'en waves as they open up the background magnetosphere \citep{Yuan:2020pea,Yuan:2022mbd}, although most of the energy of these Alfv\'en waves is spent deforming the background itself \citep{Yuan:2022mbd,Most:2024eig}.

Starting from $t = 100\,r_*/c$, the magnetosphere becomes fully disrupted due to the higher-frequency components of secondary MPs. This disruption leads to the formation of unstable and elongated equatorial current sheets, which dissipate a substantial fraction of the background magnetospheric energy through reconnection. In our computation of the various energy components, both {persistent} deformations and the dissipation of the background magnetic field are included in the energy budget of the outgoing MPs, primarily in the fast magnetosonic energy channel.

As a result, beyond $t = 100\,r_*/c$, the Alfv\'en and fast magnetosonic energies no longer represent purely wave-related energy. Instead, the observed decrease in fast magnetosonic energy primarily reflects the dissipation of background magnetospheric energy. The dissipated energy $\mathcal{E}_\mathrm{dis}$ shown in Figure \ref{fig:energyproportion} therefore accounts for both the dissipation associated with MPs and that of the background magnetosphere itself.

Once the magnetosphere is fully disrupted, the trapped Alfv\'en waves begin to leak from the closed magnetic field loops, which progressively open as the current sheets develop. A dominant source in this process is likely the self-interaction of the long continuously injected Alfv\'en wave trains on closed field lines which can lead to rapid fast mode conversion (at twice the Alfven wave frequency \citep{Chen:2024jvx}) for trains longer than the length of the closed flux tubes they propagate on \citep{Bernardi:2024upq}.

The outgoing Alfv\'en wave energy exhibits a slightly different behavior. It initially dissipates rapidly, as nonlinear Alfv\'en waves displace background magnetic field lines to propagate, effectively converting part of their energy into static deformations of the magnetosphere. At later times, the trapped Alfv\'en waves are released as the closed zone is peeled open, which accounts for the increase in outgoing Alfv\'en energy, even as the total perturbation energy continues to decrease.

After the generation of the first four nonlinear fast pulses and nonlinear Alfv\'en ejecta, i.e., around $t = 100\,r_*/c$, the higher-frequency components of the CPs start to become more important, while the main component is attenuated. We can see this by comparing the time evolution of the energy the surface motion of the crust with that injected into the magnetosphere. We also compute the frequency spectrum directly using a Fourier transform of that data (Figure \ref{fig:crustmagneto}). This is due to the propagation of CPs throughout the crust, which enhances the high-frequency content generated by out-of-phase reflections of these perturbations at the crust–magnetosphere and crust-core interfaces. The magnetosphere is thus filled with higher-frequency MPs, without any real interruption, ultimately leading to the disruption of the equatorial current sheets that still connected the inner magnetosphere to the first nonlinear Alfv\'en ejecta. It is at this stage that magnetic reconnection in the equatorial current sheets becomes strongly enhanced. The magnetosphere is no longer correctly described by a dipole magnetic field, having transitioned to a partially split monopole. This affects the dissipation of MPs during propagation, which is strongly dependent on the profile of the background magnetic field. Overall, the main frequency content of the crustal surface motion is reflected also in the magnetospheric waves, see Figure \ref{fig:crustmagneto}. In particular, we find that the lowest order frequencies dominate the energy budget of the magnetospheric waves.

In general, in a neutron star with crust thickness $H\sim 10^5$~cm and elastic wave speed $v_s \sim 10^8$~cm, the elastic wave crossing time of the crust thickness is $t_{\rm el} = H/v_{\rm el}\sim 10^{-3}$~s. The characteristic frequency of elastic waves bouncing between the crust-core interface and the surface is then $\nu_0= 1/(2t_{\rm el})\approx 500$~Hz. The precise value correlates strongly with $H$, which can change by a factor of $2\times$ depending on the mass of the neutron star \citep{Grill:2014aea}. In fact, for the system we adopt we find $\nu_{\rm simulation} \approx 1$~kHz. Overall, this frequency range is similar to  QPOs observed in a hyperactive FRB source \citep{Zhou:2025acx}.

\section{Conclusions}

We have provided the first investigation of nonlinear magnetospheric dynamics following a realistic neutron star crustal quake. This resulted in the injection of interesting transient features into the magnetosphere, including strong magnetized shocks, low-frequency fast magnetosonic pulses, trapped Alfv\'en waves, nonlinear Alfv\'en wave ejecta, equatorial reconnecting current sheets and associated plasmoids advected by the ejecta. 

While individual (and idealized) aspects of such dynamics have been considered before, our work demonstrates the highly nonlinear nature of a realistic magnetospheric event triggered by crustal activity. In particular, we find that nonlinear Alfv\'en wave ejecta and monster shocks \citep{Chen:2022yci,Beloborodov:2023mrs,Most:2024qgc,Bernardi:2025wht} form almost simultaneously, with roughly equal amounts initially contained in the injected linear Alfv\'en and fast magnetosonic waves \citep{qu2025threedimensionalnumericalsimulationsmagnetar}. Although previous work considered magnetospheric waves injected by a single sudden twist of the surface \citep{Yuan:2020pea,Yuan:2022mbd}, we find that the subsequent emission of multiple nonlinear Alfv\'en wave ejecta events rips open and combs out large part of the magnetosphere, which does not remain in its initial dipole shape. Most outgoing energy is contained in nonlinear fast magnetosonic waves, which {are expected to form } monster shocks \citep{Chen:2022yci,Beloborodov:2023mrs}. Our simulations, indeed show qualitative indications for  monster shock formation for at least the leading waves \footnote{Since monster shock formation is an MHD effect, we can only approximately capture it using a limiter on the electric fields in our FFE simulation that mimics some of the expected behavior \citep{Most:2024qgc}. In reality, we expect all kilohertz fast magnetosonic waves to form shocks \citep{Beloborodov:2023mrs}.}. In between emission events transient current sheets form, which could be the source of X-ray emission \citep{Yuan:2020pea}. The  magnetospheric waves retain the dominant periodicity of the crustal oscillations (Figure \ref{fig:crustmagneto}). Therefore, our simulation makes a clear prediction of short lived QPOs of frequency $\nu\sim 0.5-2$~kHz, depending on the thickness of the crust.

Our work, has several implications. First, pure Alfv\'en wave ejecta contain only a percent-level fraction of the crustal energy, with most of the Alfv\'en waves remaining trapped (we caution that due to the highly dynamical nature of the magnetosphere it has been difficult to assess fast magnetosonic waves conversion in trapped regions \citep{Yuan:2020eor,Bernardi:2024upq,Mahlmann:2024gui}. 

Nonlinear Alfv\'en waves convert efficiently into fast waves, but primarily deform the background, though this efficiency is one order of magnitude lower than what single Alfv\'en wave pulse have predicted \citep{Yuan:2022mbd,Most:2024eig}. {In addition to fast magnetosonic waves converted from these Alfv\'en wave ejecta, we find that} half the injected energy propagates outward in the form of strong magnetized shocks, and nonlinear magnetosonic waves.

In conjunction with energy constraints on hyperactive repeating sources \citep{Zhang:2025qzn}, this may potentially place constraints on some proposed FRB mechanisms \citep{Yuan:2020pea}, though we caution that more detailed studies both of the crust modeling and the magnetospheric modeling will be required. Second, our work shows that the background magnetosphere transitions to a strongly combed out partially split-monopole like structure, which may impact potential damping \citep{Beloborodov:2023lxl,Qu:2022got} and emission mechanism \citep{Lu:2020nsg,Long:2024wef}. Third, crustal oscillation frequencies are imprinted onto the magnetospheric dynamics. This may have implications for hyperactive repeating sources \citep{2025ApJ...979L..42W} and potential quasi-periodic oscillations \citep{Zhou:2025acx}, although part of these could also be associated with intrinsic subpulse variability in neutron star radio sources \citep{Kramer:2023mga}.

While providing a first glimpse on the complex nature of what a crustal event could look like in terms of magnetospheric dynamics, future work is required both in modeling different crustal perturbations {and different crust thicknesses}, but also in properly capturing magnetospheric dynamics using (radiation) MHD \citep{Most:2024qgc}, or kinetic approaches \citep{Bernardi:2025wht}.

\begin{acknowledgments}

The authors are grateful for discussions with Nils Andersson, Michael Grehan, E. Sterl Phinney, Yuanhong Qu, Alexis Reboul-Salze, Bart Ripperda, and Christopher Thompson.
Simulations were performed on DOE NERSC supercomputer Perlmutter under grants m4575 and m5801, which uses resources of the National Energy Research Scientific Computing Center, a DOE Office of Science User Facility supported by the Office of Science of the U.S. Department of Energy under Contract No. DE-AC02-05CH11231 using NERSC award NP-ERCAP0028480. ERM acknowledges support by the National Science Foundation under grants No. AST-2307394, from NASA's ATP program under grant 80NSSC24K1229, as well as on the NSF Frontera supercomputer under grant AST21006, and on Delta at the National Center for Supercomputing Applications (NCSA) through allocation PHY210074 from the Advanced Cyberinfrastructure Coordination Ecosystem: Services \& Support (ACCESS) program, which is supported by National Science Foundation grants \#2138259, \#2138286, \#2138307, \#2137603, and \#2138296. A.B. is supported by a PCTS fellowship and a Lyman Spitzer Jr. fellowship.

\end{acknowledgments}

\software{
      AMReX \citep{amrex},
	  matplotlib \citep{Hunter:2007},
	  numpy \citep{harris2020array},
	  scipy \citep{2020SciPy-NMeth},
      yt \citep{2011ApJS..192....9T}.
}

\bibliography{louis, elias, software}{}
\bibliographystyle{aasjournalv7}

\end{document}